# Nonvolatile optical control of interlayer stacking order in 1$T$-TaS$_2$

Junde Liu[1,†], Pei Liu[2,3,†], Liu Yang[2,3,†], Sung-Hoon Lee[4], Mojun Pan[1], Famin Chen[1], Jierui Huang[1], Bei Jiang[1], Mingzhe Hu[1], Yuchong Zhang[1], Zhaoyang Xie[2,3], Gang Wang[2,3], Mengxue Guan[2,3], Wei Jiang[2,3], Huaixin Yang[1], Jianqi Li[1], Chenxia Yun[1], Zhiwei Wang[2,3], Sheng Meng[1], Yugui Yao[2,3,*], Tian Qian[1,*], Xun Shi[2,3,*]

[1]Beijing National Laboratory for Condensed Matter Physics and Institute of Physics, Chinese Academy of Sciences, Beijing, China

[2]Centre for Quantum Physics, Key Laboratory of Advanced Optoelectronic Quantum Architecture and Measurement (MOE), School of Physics, Beijing Institute of Technology, Beijing, China

[3]Beijing Key Lab of Nanophotonics and Ultrafine Optoelectronic Systems, Beijing Institute of Technology, Beijing, China

[4]Department of Applied Physics, Kyung Hee University, Yongin, Republic of Korea

[†]These authors contributed equally to this work.

[*]Corresponding author. Email: ygyao@bit.edu.cn; tqian@iphy.ac.cn; shixun@bit.edu.cn

**Nonvolatile optical manipulation of material properties on demand is a highly sought-after feature in the advancement of future optoelectronic applications. While the discovery of such metastable transition in various materials holds good promise for achieving this goal, their practical implementation is still in the nascent stage. Here, we unravel the nature of the ultrafast laser-induced hidden state in 1$T$-TaS$_2$ by systematically characterizing the electronic structure evolution throughout the reversible transition cycle. We identify it as a mixed-stacking state involving two similarly low-energy interlayer orders, which is manifested as the charge density wave phase disruption. Furthermore, our comparative experiments utilizing the single-pulse writing, pulse-train erasing and pulse-pair control explicitly reveal the distinct mechanism of the bidirectional transformations — the ultrafast formation of the hidden state is initiated by a coherent phonon which triggers a competition of interlayer stacking orders, while its recovery to the initial state is governed by the progressive domain coarsening. Our work highlights the deterministic role of the competing interlayer orders in the nonvolatile phase transition in the layered material 1$T$-TaS$_2$, and promises the coherent control of the phase transition and switching speed. More importantly, these results establish all-optical engineering of stacking orders in low-dimensional materials as a viable strategy for achieving desirable nonvolatile electronic devices.**

Low-dimensional materials exhibit flexible tunability in physical properties, owing to the reduced dimensionality and the interlayer/interchain van der Waals coupling. A subtle modification of lattice arrangement could significantly alter the electronic behaviours of these materials, making them ideal for studying structure-function relations and searching for desirable functionalities [1–6]. Among the numerous tools available for such school of engineering, light stands out as a unique one due to its versatility, non-contact nature and potentially ultrafast speed [7–16]. Ultrafast laser pulses have the capability to drive materials far from thermodynamic equilibrium. Subsequently, the system typically relaxes to its initial state over a characteristic timescale, however, in certain instances, it stabilizes into a laser-induced new state [7,8,17,18]. This intriguing phenomenon has been discovered, sometimes serendipitously, in several materials spanning a diverse class [19–21]. The nonvolatility of these transitions significantly enhances their feasibility for applications in optoelectronics, particularly high-performance optical storage and switching. Yet, the practical exploitation of this potential [22,23] remains limited by the current lack of sufficient control and optimization of such transitions, highlighting the urgent need for more understanding of the underlying light-matter interactions.

As an exemplary case of the photoinduced nonvolatile transition, the metallic hidden state in $1T$-$TaS_2$ has been widely studied since its discovery [19,24–33]. The new state can be formed and stabilized upon a single intense laser excitation, featuring a prominent reduction in resistivity by orders of magnitude. Under thermal equilibrium, this layered chalcogenide hosts a complex phase diagram including multiple charge density wave (CDW) orders [34,35]. The commensurate CDW (C-CDW) order developed at the temperature ~180 K is characterized by an intralayer lattice distortion with the formation of star-of-David (SOD) clusters (Fig. 1(b)), and is accompanied by interlayer CDW stacking with layer dimerization [36–45] (Fig. 1(c)). After being transformed into the hidden state, previous studies have revealed the formation of domains and domain walls, which disrupt the in-plane CDW phase coherence [30,46]. Moreover, the X-ray diffraction suggested that this transition also involves a collapse of layer dimerization in the out-of-plane direction [29]. Compared to the structural characterization, the reported electronic information of the hidden state suggests a metallic nature, while its exact nature is rarely explored. A detailed investigation into the band structure and its evolution during the switch remains elusive, hindering a comprehensive understanding of the nature of the hidden state, and more broadly, the strategic search for nonvolatile phase transitions.

In this work, we utilize a single intense femtosecond laser pulse to steer 1*T*-TaS$_2$ into the hidden state, which remains stable until it is reverted back to the C-CDW state by using a weak pulse train, as illustrated in Fig. 1. During this repeatable cycle, the electronic band structure is monitored *in situ* by angle-resolved photoemission spectroscopy (ARPES). The spectrum of the hidden state is intimately connected to that of C-CDW, but remarkably, there emerge additional features that are consistent with a new similar low-energy interlayer configuration (Fig. 1(d)), suggesting the formation of a mosaic phase with mixed stacking orders (Fig. 1(e)). The relative volume of such mixture is closely related to the macroscopic metallicity, and can be progressively tuned by the erasing pulse sequence. Moreover, the pulse-pair control experiments reveal that the domain formation during the transition from the C-CDW to the hidden state is triggered by the strongly excited coherent CDW amplitude mode, which smears out the diversity of stacking and induces a competition between low-energy orders. Such a coherent process of collective atomic motion is ultrafast with the timescale of sub-picosecond. In contrast, the reverse transition from the hidden to C-CDW state is dominated by the incoherent domain coarsening. Our results further elucidate the microscopic nature of the hidden state and bidirectional switch in 1*T*-TaS$_2$, demonstrating that the competing interlayer orders and coherent excitation can be harnessed to actively explore more optimized optical control of low dimensional materials.

We first investigate the low-energy electronic structure of 1*T*-TaS$_2$ in both the C-CDW and hidden states, and compare them to theoretical calculations [45], as depicted in Fig. 2. Specifically, Figs. 2(c) and 2(g) display the band structures of the initial C-CDW state along the $\bar{\Gamma} - \bar{M}$ direction, as recorded with the He I$\alpha$ (21.2 eV) and the laser (7.2 eV) sources, respectively, clearly showing the insulating behaviour and multiple CDW gaps. These results align closely with the calculated band structure for the AL stacking configuration as defined in Fig. 1(c), indicating a layer dimerization in the C-CDW state that is consistent with previous reports [37–39,42]. Following the photoexcitation by a single intense laser pulse at 1.2 eV with the fluence of ~ 1.2 mJ/cm$^2$ and the pulse duration of ~ 80 fs, we successfully induced a transition from the C-CDW to the hidden state. The ARPES intensity map (Figs. 2(e) and 2(i)) of this hidden state measured in the same way, as well as the corresponding curve analysis of the density of states (DOS, Fig. 2(l)) and energy distribution curves (EDCs, Fig. 2(m)), clearly reflect its metallic nature. A closer investigation into the band dispersion and spectral weight uncovers the appearance of additional features upon the transformation from the C-CDW to the hidden state, including the CDW gap filling (not closing) and a band crossing the Fermi level ($E_F$). This is distinct from a complete modification of the electronic structure and indicates a coexistence of different orders within the probe area, which is reminiscent of the domain formation as observed by the scanning tunnelling microscopy (STM) [30,47–50]. Note that the domains were structurally indistinguishable in the STM characterizations of the top layer, the domain walls were initially considered as the conducting channels. However, this scenario is inconsistent with the scanning tunnelling spectroscopy results, and the metallic nature of the hidden state remains to be understood [30,35,47,49].

Since the intralayer SOD structure is retained in the hidden state, we consider the possible variation of the interlayer stacking. As shown in Fig. 2(k), the stability evaluation of different interlayer configurations reveals that the L stacking (defined in Fig. 1(d)) is very close to the ground-state AL stacking in terms of the total energy, and meanwhile, they are notably lower than the others [45]. Indeed, the mixed spectra of AL and L stackings, rather than one of them entirely, reproduce well the experimental observations, with the key features highlighted by the arrows in Figs. 2(e)-(f) and 2(i)-(j) (see also Fig. S10). Such consistency uncovers the coexistence of the insulating AL and metallic L configurations in the hidden state, and thus clarifies the debatable origin of the metallicity [30,35,47,49,51]. This mixing is a coupled in-plane and out-of-plane reconstruction effect governed by the spatial discontinuity of the relative SOD phase between neighbouring layers, manifested as the appearance of domains [30] and the collapse of long-range layer dimerization [29,45]. Despite the small energy difference between these two interlayer orders, the domain wall formed during the laser-induced nonequilibrium process could provide a sufficient energy barrier to stabilize this hidden state at temperatures below ~ 46 K (Fig. S1).

In order to gain more insights into this mosaic hidden state, we investigate how it can be reversed back to the C-CDW state. As any excitation with a strong laser pulse above a threshold of ~ 0.8 mJ/cm$^2$ (Figs. S4 and S5) would introduce the domain formation, we utilize a relatively weak pulse sequence (photon energy = 1.2 eV, fluence ~ 0.48 mJ/cm$^2$, pulse number = 500, pulse duration ~ 80 fs) to erase the hidden state. As shown in Fig. 3, a series of such pulse sequences induce a continuous transition towards the C-CDW state, until a complete erasure is achieved (see the remarkable similarity between the erased state and the initial C-CDW state in Fig. 3(a)). By tracing the ARPES intensity within the marked region in Fig. 3(a), where the new emerging band crosses $E_F$, we quantitatively plot this progressive evolution in Fig. 3(c). The result reveals an exponential-like suppression of the electronic features associated with the L stacking, implying a coarsening of the AL domains upon the sequential laser annealing, and eventually forming the C-CDW state. The recently discussed surface dimerization or reconstruction may also facilitate this process [41]. In addition, a systematic comparison between the measured dispersions to the calculations uncovers an increasing AL to L ratio during the erasing process (see supplemental material Section 6), thus further validating the scenario of stacking coexistence. It is also worth mentioning that such a progressive evolution in 1$T$-TaS$_2$ enables a continuous and controllable optical tuning of its resistivity spanning three orders of magnitude.

Having identified the nature of the hidden state, we next investigate the mechanism of this nonvolatile transition. Motivated by the coherent control of a structural transition [52], we replace the single writing (*i.e.*, C-CDW to hidden state) pulse by a pulse pair with varying time delay (illustrated in Fig. 4(a)), to check its effect on the transition as well as the possible coherence. Following the same practice described above, we select the region covering the metallic band in the hidden state, as indicated in Figs. 4(c)-(d), to characterize the writing efficiency. As shown in Fig. 4(e), the intensity of this region exhibits a general trend of efficiency decrease as the two pulses moving away from each other, but at the same time, superimposed by a strong modulation. Fitting to this oscillatory behaviour gives a frequency of ~ 2.21 THz, which corresponds to the CDW amplitude mode [19,27,53], pointing out its intimate relation to the hidden state formation. In fact, a strong excitation of this breathing mode of the SOD could suppress and even reverse the periodic lattice distortion [54], leading to a large modulation in the energy landscape/hierarchy of different interlayer stackings shown in Fig. 2(k). In the case of the lattice being transiently restored to the normal state, all these stacking orders defined by the alignment of Ta atoms in neighbouring layers become identical. Such a nonequilibrium condition appears only under an above-threshold excitation, triggering the possible competition and reformation of interlayer stackings. Therefore, the slip of the SOD phase (*i.e.*, centring at a different Ta site) during the following rapid relaxation would introduce multiple stackings, and being stabilized to the above-mentioned AL and L domains considering their comparable low energies. As a comparison, the erasing process shows no signs of coherent response to the dual-pulse control (Fig. 4(f)), which is primarily attributed to the spatial incoherence of the hidden state with mixed stacking configurations. This scenario is illustrated in Fig. 1(e), highlighting the dichotomy of the bidirectional optical switch. In this sense, the speed of the coherent writing process from C-CDW to the hidden state is estimated to be at terahertz, while the erasing process involving the domain dynamics might be much slower.

Based on the comprehensive results, we shall summarize the key factors enabling the nonvolatile optical manipulation of $1T$-TaS$_2$. First, a delicate balance of the atomic structure developed in the CDW state makes a unique energy hierarchy that contains multiple interlayer configurations near the ground state [41,45] (*i.e.*, AL vs L here). A proper external perturbation in the vicinity of such stacking competition drives the material to a hidden state that can be stabilized, thus assuring the nonvolatility of the transition. Second, the competing interlayer orders are distinguished by a phase shift of CDW lattice distortion, while nurturing distinct electronic structures. As such, this transition induces a remarkable modification of material properties without involving a large lattice reorganization, potentially supporting an energy-efficient switch. Third, the photoexcited CDW amplitude mode can smear out the distinction of these stacking orders and coherently trigger their competition, pushing the speed of the optical transition towards the limit of the intrinsic phonon frequency.

Our work clarifies the nature and microscopic origin of the hidden metallic state in $1T$-TaS$_2$, highlighting the central role of the interlayer stacking competition and its coherent excitation. More importantly, our results and understandings demonstrate a feasible strategy for inducing nonvolatile photo control in more low-dimensional materials beyond $1T$-TaS$_2$. Specifically, one can explore the materials containing nearly competing interlayer/interchain orders, or create such competition via chemical decorations, etc. On this basis, a coherent and properly controlled excitation of the certain phonon mode associated with the transition between competing orders, could possibly induce an all-optical reversible phase transition. Future research along this line appears to be a promising route to eventually achieve the desirable photoswitch that is ultrafast, nonvolatile, reversible, and operating at room temperature.

**Acknowledgement:** We thank H. Ding for insightful discussions. This work was supported by the National Key Research and Development Program of China (grants nos. 2020YFA0308800 and 2022YFA1403400), the National Natural Science Foundation of China (grant nos. 12204042, 12025407, and U2032204), the Ministry of Science and Technology of China (grant no. 2022YFA1403800), the Beijing Natural Science Foundation (grant no. Z210006), the Beijing Institute of Technology (BIT) Research Fund Program for Young Scholars, the Chinese Academy of Sciences (grant no. XDB33000000), the Beijing National Laboratory for Condensed Matter Physics (grant no. 2023BNLCMPKF007), and the Synergetic Extreme Condition User Facility. Z.W. thanks the Analysis and Testing Center at BIT for assistance in facility support. S.-H.L. was supported by the National Research Foundation of Korea (grant no. 2018R1A2B6004044).

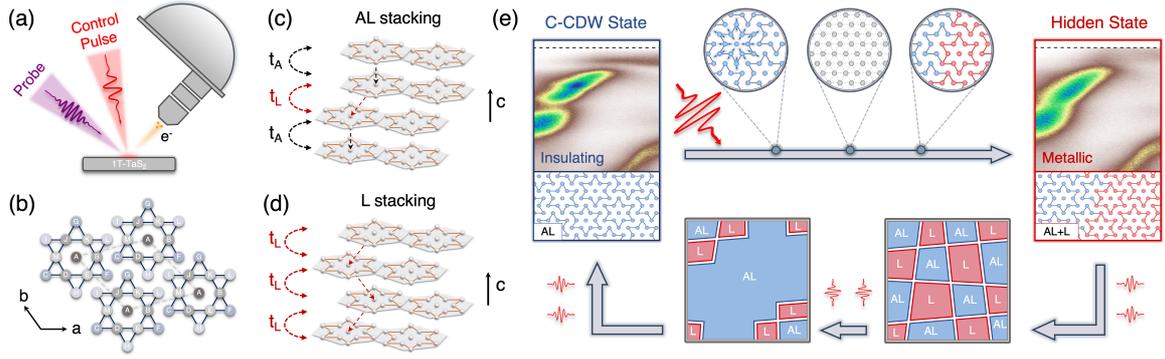

FIG. 1. Bidirectional optical manipulation of the material state governed by stacking order in $1T$-TaS$_2$. (a), Schematic illustration of the ARPES measurements under different phases controlled by laser pulses. (b), Illustration of the in-plane star-of-David cluster in the C-CDW state, with labelling conventions for the 13 Ta atoms in each star. Dashed lines indicate the superlattice structure. (c), (d), Illustration of AL stacking and L stacking arrangements. The black dashed line indicates the out-of-plane hopping ($t_A$) where A-atoms are vertically aligned. The red dashed line indicates the out-of-plane hopping ($t_L$) where A-atom and L-atom are randomly aligned. (e), Schematic illustration of bidirectional transitions between the insulating C-CDW state (AL stacking) and the metallic hidden state (AL + L stacking). The upper panel illustrates the writing process induced by a single pulse, via the excitation of coherent amplitude mode, the expansion to a symmetric state and eventually the formation of a hidden domain state. The lower panel illustrates the continuous erasing process implemented by the pulse train where the domain of L stacking gradually diminishes.

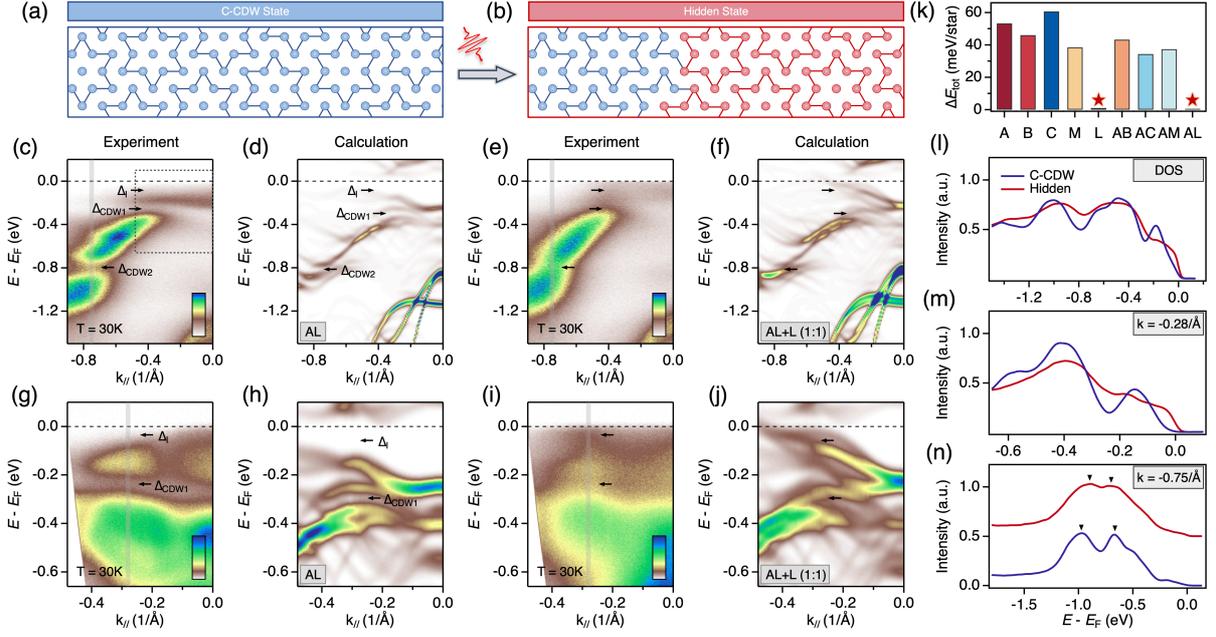

FIG. 2. Experimental and theoretical results of the C-CDW state and hidden state. (a), (b), Schematic illustration of the transition from the C-CDW state to the hidden state induced by a single laser pulse with a fluence of ~ 1.2 mJ/cm$^2$. (c), Experimental band structure of the C-CDW state along the $\bar{\Gamma} - \bar{M}$ direction measured by a helium lamp source (21.2 eV) at T = 30 K. (d), Calculated band structure of the C-CDW state along the $\bar{\Gamma} - \bar{M}$ direction [45]. (e), (f), Similar to panels (c) and (d) but for the hidden state. (g)-(j), Similar to panels (c)-(f) but recorded with a laser source (7.2 eV), and for a zoom-in region indicated by a dashed rectangle in panel (c). The black arrows highlight the distinct spectral features in two states. (k), Total energy of the system plotted against different interlayer stacking orders. The lowest-energy AL stacking and the second-lowest L stacking are highlighted with red pentagrams. (l), DOS for the C-CDW state (blue curve) and the hidden state (red curve) in panels (c) and (e), respectively. (m), EDCs for the C-CDW state (red curve) and the hidden State (blue curve) in the region marked by a grey line in panels (g) and (i), respectively. (n), EDCs for the C-CDW state (blue curve) and the hidden State (red curve) in the region marked by a grey line in panels (c) and (e), respectively. The arrows indicate the positions of the peaks.

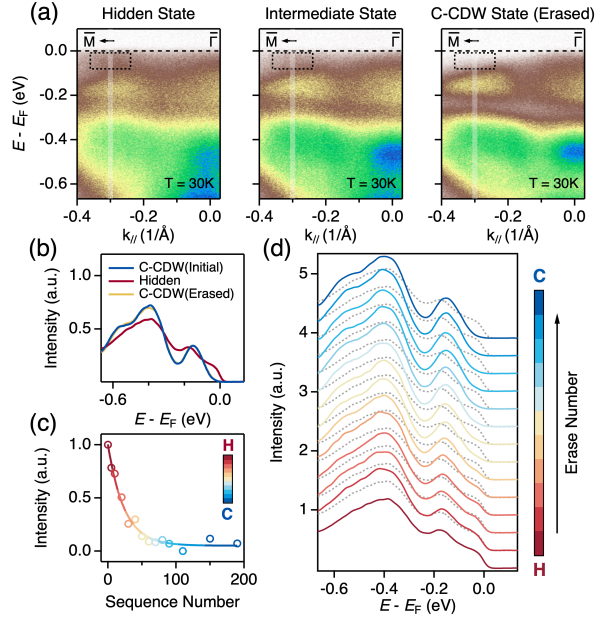

FIG. 3. Continuous erasing process from the hidden state to the C-CDW state. (a), Band structure of the hidden state, intermediate state after partial erasing, and the C-CDW state after complete erasing along the $\bar{\Gamma} - \bar{M}$ direction measured by a laser source (7.2 eV) at T = 30 K, respectively. (b), EDCs of the initial C-CDW state (blue curve), the hidden state (red curve) and the C-CDW state after complete erasing (yellow curve) in the region marked by a grey line in panel (a), respectively. (c), Spectral intensity integrated within the region marked by a dashed rectangle in panel (a) as a function of erasing sequence number. (d), EDCs in the region marked by a grey line in panel (a) under different intermediate states (from the hidden state to the C-CDW state) with increasing erasing sequence number. Dashed lines indicate the EDCs of the hidden state for comparison.

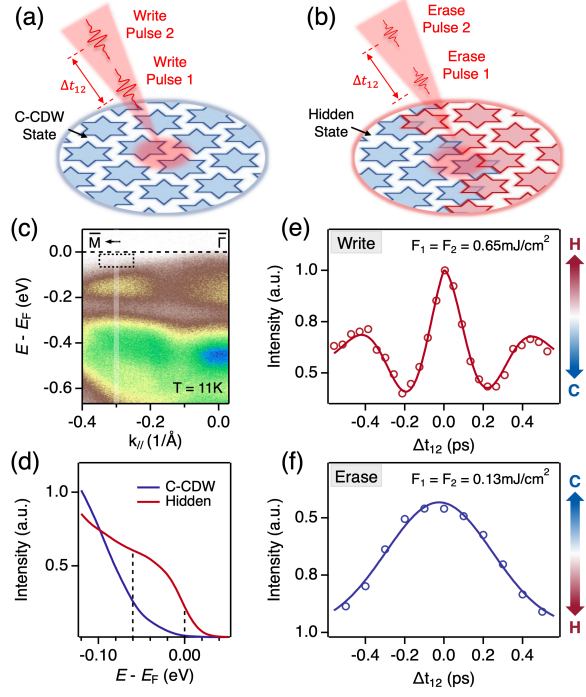

FIG. 4. Dual-pulse control of the writing and erasing processes. (a), (b), Schematic illustration of the dual-pulse writing process and erasing process, respectively. The C-CDW state corresponds to the AL stacking (blue stars), and the hidden state corresponds to a mixture of AL stacking (blue stars) and L stacking (red stars). The $\Delta t_{12}$ represents the time delay between two laser pulses. (c), Band structure of the C-CDW state along the $\bar{\Gamma} - \bar{M}$ direction measured by a laser source (7.2 eV) at T = 11 K. (d), EDCs of the C-CDW state (blue curve) and the hidden state (red curve) in the region marked by a grey line in panel (c), respectively. The hidden state is achieved by complete dual-pulse writing with $\Delta t_{12} = 0$. (e), ARPES intensity integrated within the region marked by a dashed rectangle in panel (c) as a function of the time delay $\Delta t_{12}$. (f), Similar to panel (e) for the dual-pulse erasing process.

# Supplemental Material for
# Nonvolatile optical control of interlayer stacking order in 1$T$-TaS$_2$

**This PDF file includes:**

**Section S1. Laser excitation and ARPES measurements.**

A Yb-doped fiber laser system generated the control laser pulse with a photon energy of 1.2 eV, a pulse duration of ~ 80 fs and a spot size of ~ 100 um. The repetition rate of the laser could be adjusted between 1 Hz and 1 MHz. To enable the selection of a single pulse or pulse sequences for the writing and erasing process, an optical shutter synchronized with the laser output signal was employed. ARPES measurements were performed in ultra-high vacuum better than $3 \times 10^{-11}$ mbar, using a hemispherical electron analyser (Scienta Omicron DA30-L) equipped with a helium discharge lamp source and a laser source [1]. The photon energy of He Iα from the helium lamp is 21.2 eV with the overall energy resolution better than 5 meV and the spot size of ~1 mm. The photon energy of the vacuum-ultraviolet laser is 7.2 eV with the overall energy resolution of ~ 10 meV and the spot size of ~ 20 um. Sample manipulation was facilitated by a 6-axis manipulator with temperature control ranging from 6 to 300 K. Samples measured at 30 K or 11 K were cleaved *in situ* at the respective temperature.

**Section S2. Density functional theory (DFT) calculations.**

We conducted our DFT study using the Vienna Ab Initio Simulation Package (VASP) [2–4] with the generalized gradient approximation (GGA) [5]. To account for van der Waals interactions, we used the Tkatscenko-Scheffler scheme [6]. We expanded the electron wave functions using a plane wave basis set with an energy cutoff of 260 eV. K-point integration was carried out using a uniform k-point mesh with a density of $4 \times 4 \times 8$ in the Brillouin zone of the $\sqrt{13} \times \sqrt{13} \times 1$ unit cell. To achieve relaxation, all atoms were allowed to move until the residual forces reached below 0.02 eV/Å. Our calculations for the insulating phase with AL CDW stacking yielded equilibrium lattice constants of $a = 3.35$ Å and $c = 5.79$ Å, which closely matched experimental values of $a = 3.35$ Å and $c = 5.79$ Å [7].

**Section S3. Erasure of the hidden state via thermal cycling.**

In addition to the laser pulse sequences described in the main text, heating the sample presents an alternative method for erasing the hidden state. Analysis of the band dispersions in Figs. S1(a)-(c) and the EDCs in Fig. S1(d) confirms that a thermal cycling effectively reverts the hidden state to the C-CDW state, manifested as the re-establishment of the CDW gap and a transition back to insulating behaviour. The thermal stability of the hidden state was further investigated through the temperature-dependent evolution of the EDCs near the CDW gap (Fig. S1(e)), revealing a gap opening with increasing temperature. Additionally, as illustrated in Fig.

S1(f), fitting the peak position at $E_B \sim 0.7$ eV against temperature determines a characteristic transition temperature of ~ 46 K.

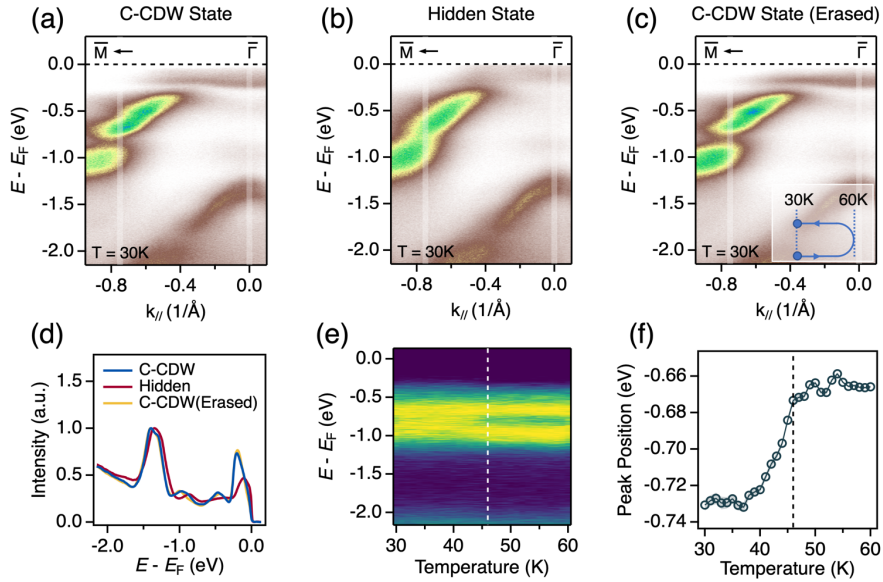

FIG. S1. Thermal cycling erasing of the hidden state. (a)-(c), Band structures along the $\bar{\Gamma} - \bar{M}$ direction for the initial C-CDW state, the hidden state and the C-CDW state after thermal cycling erasing, measured by a helium lamp source (21.2 eV) at $T = 30$ K, respectively. (d), EDCs for the initial C-CDW state (blue curve), the hidden state (red curve) and the C-CDW state after erasing (yellow curve) in the region marked by the grey line at the $\bar{\Gamma}$ point in panels (a)-(c), respectively. (e), EDCs in the region marked by the grey line near the $\bar{M}$ point in panel (b) as a function of temperature. (f), Corresponding positions of the peaks at $E_B \sim 0.7$ eV as a function of temperature.

### Section S4. Suppression of the long-range interlayer stacking order and $k_z$ broadening.

Recent research suggests that the extensive bandwidth in the C-CDW state originates from the layer dimerization and strong interlayer hopping, implying that the C-CDW state may be classified as a band insulator [7–15] (Fig. S3(a)). An intermediate state (I state), observed during the transition from the C-CDW state to the nearly commensurate CDW (NC-CDW) state, exhibits a significantly reduced bandwidth, suggesting its potential classification as a true Mott insulator [7,8] (Fig. S3(b)). Figures S2(a)-(f) depict the temperature-dependent evolution of band structures along the $\bar{\Gamma} - \bar{M}$ direction with samples cleaved at 250 K. Figures S2(g) and S2(h) present a comparison of the zoom-in dispersion images for both the C-CDW state and the I state. The comparison indicates that the band near $E_F$ exhibits a wide bandwidth and $k_z$ broadening in the C-CDW state, while the bandwidth and $k_z$ broadening of the I state are significantly reduced. This result can be attributed to the c-axis stacking disorder arising from the metallic bulk and remaining insulating surface layers in the I state, resulting in a reduction in the out-of-plane dispersion and an enhancement in the in-plane Coulomb repulsion [16].

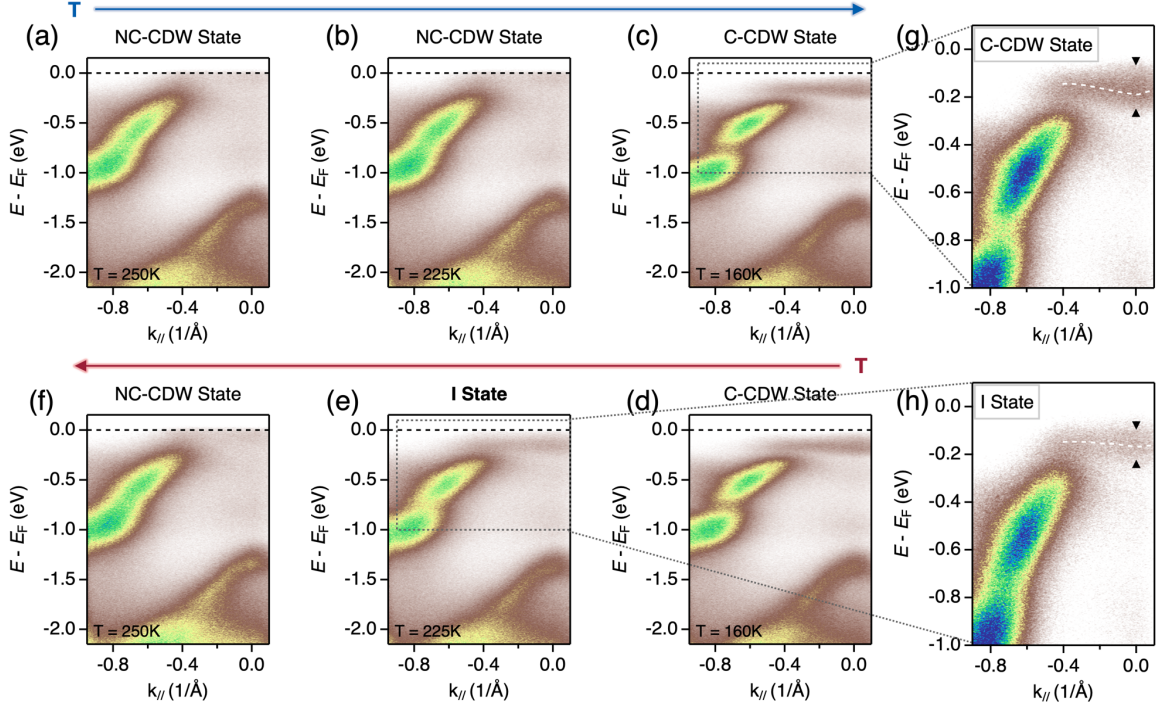

FIG. S2. Temperature-dependent evolution of the band structures. (a)-(c), Temperature-dependent band structures measured along the $\bar{\Gamma} - \bar{M}$ direction using a helium lamp source (21.2 eV) during the cooling process. (d)-(f), Similar to panels (a)-(c) but for the heating process. (g), Zoom-in dispersion marked by a dashed rectangle in panel (c) of the C-CDW state at T = 160K. (h), Zoom-in dispersion marked by a dashed rectangle in panel (e) of the I state at T = 225K.

In addition to the reduction of interlayer order in the I state, which emerges during the heating cycle under equilibrium conditions, our experimental findings suggest that optical manipulation can effectively modulate the stacking structure, thereby suppressing the long-range interlayer stacking order and $k_z$ broadening. As depicted in Fig. S3(c) and S3(d), the band near $E_F$ in the hidden state exhibits a narrower bandwidth, and the distribution of $k_z$ broadening is reduced compared to the initial C-CDW state. Further insights into these changes can be gained from the second derivative plots in Fig. S3(e) and S3(f), as well as from the comparison of the dispersions in Fig. S3(h). This result is attributed to the suppression of the interlayer order in the hidden state. As discussed in the main text, the C-CDW state exhibits layer dimerization (AL stacking), resulting in a wide bandwidth and $k_z$ broadening shown in Fig. S3(i). In contrast, the interlayer order in the hidden state is suppressed owing to the formation of domains (a mixture of AL stacking and L stacking), which effectively results in a narrower bandwidth and reduced $k_z$ broadening shown in Fig. S3(j). Additionally, the formation of the hidden state accompanied by reduced interlayer coupling may enhance the in-plane electronic correlations. Therefore, our results provide a promising platform for manipulating the ordering structure of non-equilibrium states through optical methods, thereby controlling the changes in strongly correlated electronic orders.

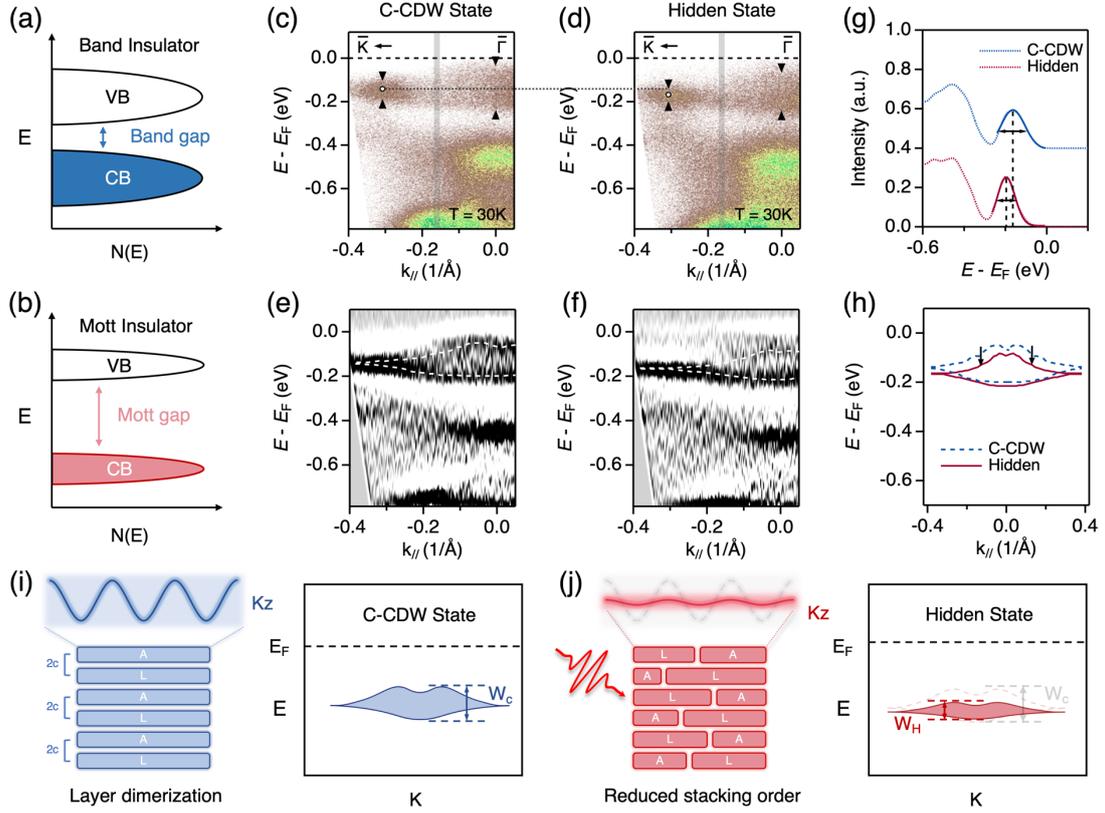

FIG. S3. Suppression of the interlayer stacking order and reduction of the $k_z$ broadening. (a), (b), Schematic illustration of the energy diagram of the band insulator and the Mott insulator, respectively. (c), (d), Band structures along the $\bar{\Gamma} - \bar{K}$ direction for the C-CDW state and the hidden state measured by a laser source (7.2 eV) at T = 30 K, respectively. The fluence of writing pulse is ~0.9mJ/cm$^2$. (e), (f), Corresponding second derivative plots of panels (c) and (d), respectively. (g), EDCs of the C-CDW state (blue curve) and the hidden state (red curve) in the region marked by a grey line in panels (c) and (d), respectively. (h), Dispersion of the low energy band for the C-CDW state (blue curve) and the hidden state (red curve). (i), Left: illustration of the $k_z$ broadening and layer dimerization in the C-CDW state. Right: illustration of the energy diagram of the C-CDW state with $k_z$ broadening. (j), Similar to panel (i) but for the hidden state with reduced $k_z$ broadening and modified stacking order.

**Section S5. More details of the optical manipulation.**

As depicted in Figs. S4 and S5, with an increase in the fluence of the writing laser, the bandwidth of the low energy band narrows, and the band signal associated with L stacking becomes more pronounced. This implies that more L stacking progressively suppresses the long-range interlayer order, thereby reducing the $k_z$ broadening. Further quantitative analysis reveals that the threshold fluence for the transition to the hidden state is ~ 0.8mJ/cm² and saturation fluence is ~ 1.2 mJ/cm².

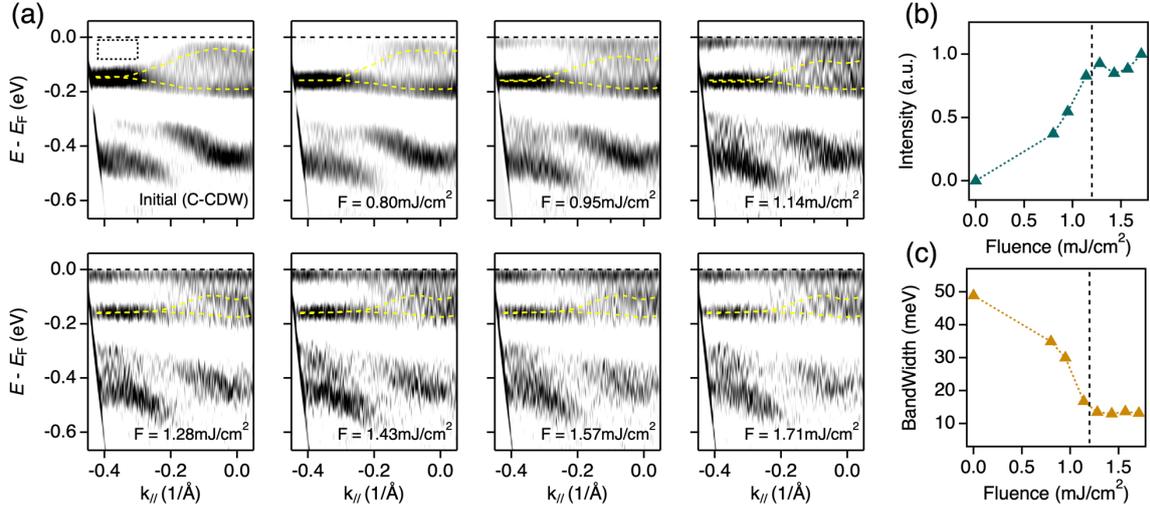

FIG. S4. Fluence-dependent dispersions along the $\bar{\Gamma} - \bar{K}$ direction. (a), Second derivative plots measured by a laser source (7.2 eV) at T = 30 K under different fluence. (b), Intensity of the region marked by a dashed rectangle in panel (a) as a function of fluence. (c), The bandwidth of the band at $E_B$ ~ 0.15eV as a function of fluence.

Furthermore, Fig. S6 illustrates the relationship between the transition to the hidden state and the number of shots from the writing laser. It is evident that the efficiency of the hidden state formation is primarily related to the fluence, while it shows little correlation with the number of shots.

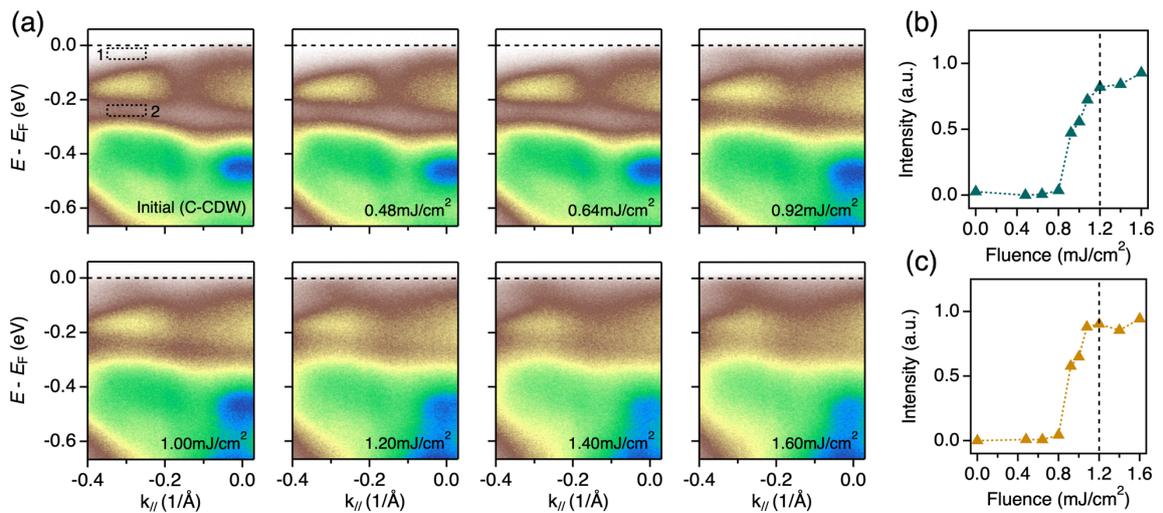

FIG. S5. Fluence-dependent dispersions along the $\bar{\Gamma} - \bar{M}$ direction. (a), Band structures measured by a laser source (7.2 eV) at T = 30 K under different fluence. (b), Intensity of the region marked by the dashed rectangle 1 in panel (a) as a function of fluence. (c), Similar to panel (b) for the region marked by the dashed rectangle 2 in panel (a).

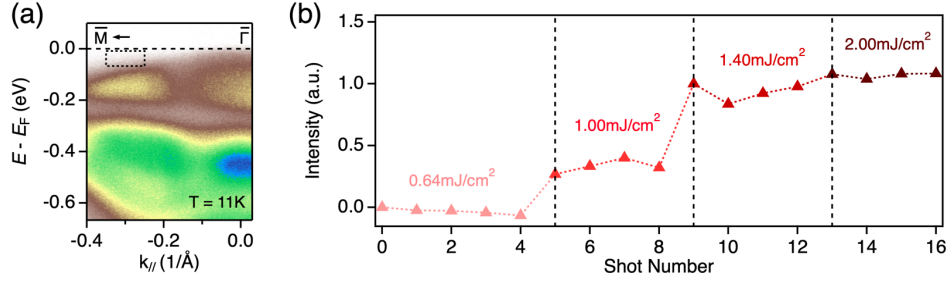

FIG. S6. Fluence-dependent evolution with an increasing shot number of writing pulse. (a), Band structures along the $\bar{\Gamma} - \bar{M}$ direction for the C-CDW state measured by a laser source (7.2 eV) at T = 11 K. (b), Shot-to-shot intensity change of the region marked by a dashed rectangle in panel (a) with an increasing fluence.

Additionally, as depicted in Fig. S7, we have investigated the variation in writing efficiency relative to the direction of the linear polarization of the writing laser. Employing the same methodology as described above, we selected the region covering the metallic band in the hidden state, as shown in Fig. S7(b), to characterize the writing efficiency. Initially, we measured the change in writing efficiency with the fluence of the writing laser with p-polarization as indicated in Fig. S7(c). Subsequently, we selected a moderate fluence with 1mJ/cm$^2$ and rotated the half-wave plate to characterize the change in writing efficiency with respect to the polarization shown in Fig. S7(d). Notably, s-polarization demonstrated higher writing efficiency compared to p-polarization. This disparity could be attributed to variations in the absorption rates and the optical transitions with the writing laser in different polarizations.

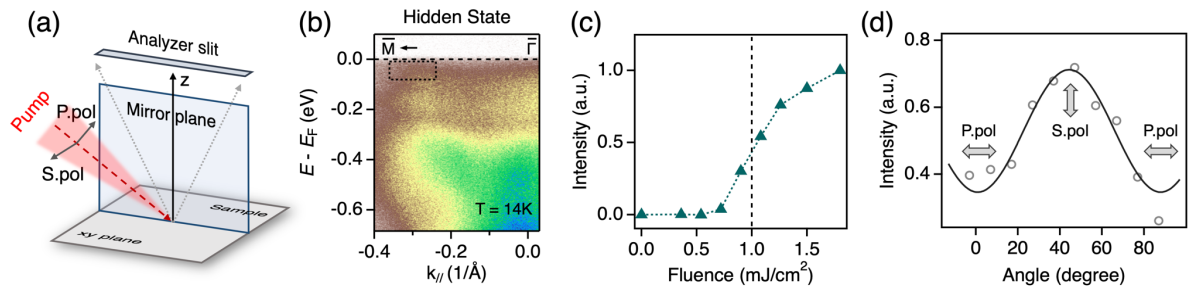

FIG. S7. Polarization-dependence of the writing efficiency. (a), Schematic experimental geometry of the laser excitation and photoemission. (b), Band structures along the $\bar{\Gamma} - \bar{M}$ direction for the hidden state measured by a laser source (7.2 eV) at T = 14 K, the writing efficiency is characterized by the intensity of the region marked by a dashed rectangle. (c), The fluence-dependent writing efficiency with p-polarization. (d), The dependence of the writing efficiency on the rotating angle of the half-wave plate.

To validate the repeatability of the bidirectional manipulation, we utilized optical methods for both the writing and erasing process, as illustrated in Fig. S8. The results indicate that the system can efficiently transition between the C-CDW state and the hidden state, demonstrating the capability of the ultrafast laser to effectively control the interlayer stacking order in this system in a bidirectional way.

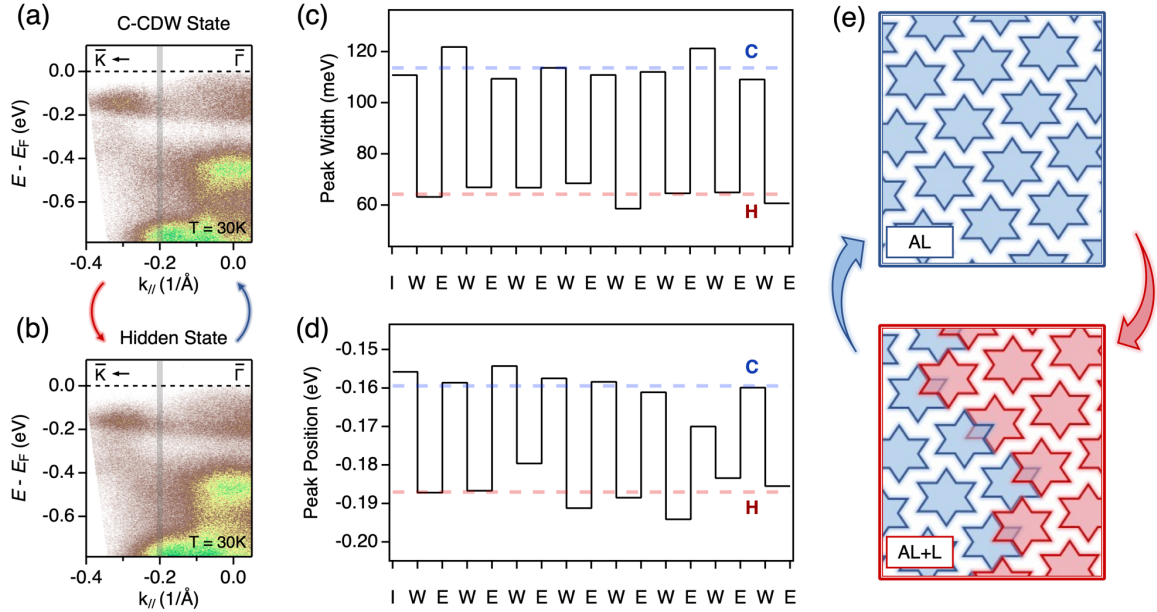

FIG. S8. Repetitive bidirectional manipulation of the states. (a), (b), Band structures along the $\bar{\Gamma} - \bar{K}$ direction for the C-CDW state and the hidden state measured by a laser source (7.2 eV) at T = 30 K, respectively. (c), The width of peaks at $E_B \sim 0.15$ eV of the EDC marked by the grey line in panels (a) and (b) during a sequence of alternating writing and erasing process. (d), Similar to panel (c) but for the positions of peaks. (e), Illustration of the bidirectional manipulation of the stacking order.

**Section S6. Comparison between the experimental and calculated band structures.**

To improve clarity regarding the correspondence of the C-CDW state to AL stacking and the hidden state being a combination of AL stacking and L stacking, we present theoretical calculation results for two interlayer configurations in Fig. S9. A comparison of the dispersions in Fig. S9(a) and S9(b) reveals the collapse of the CDW gap and the formation of a band across the Fermi level in the AL + L stacking configuration. In Figs. S9(c) and S9(d), the DOS for both configurations are illustrated, highlighting the finite intensity at the Fermi level for the AL + L stacking configuration, which is consistent with the metallic behaviour observed in experimental results. Figures S9(e) and S9(f) show the EDCs around the CDW gap, revealing the collapse of the CDW gap in the AL + L stacking configuration, which aligns with the measured results. Furthermore, EDCs in Figs. S9(g) and S9(h) suggest that the emergence of metallic behaviour can be attributed to the presence of L stacking in the hidden state. Therefore, the incorporation of L stacking in the theoretical calculation is crucial for understanding the experimentally observed metallic behaviour and the collapse of the CDW gap in the hidden state.

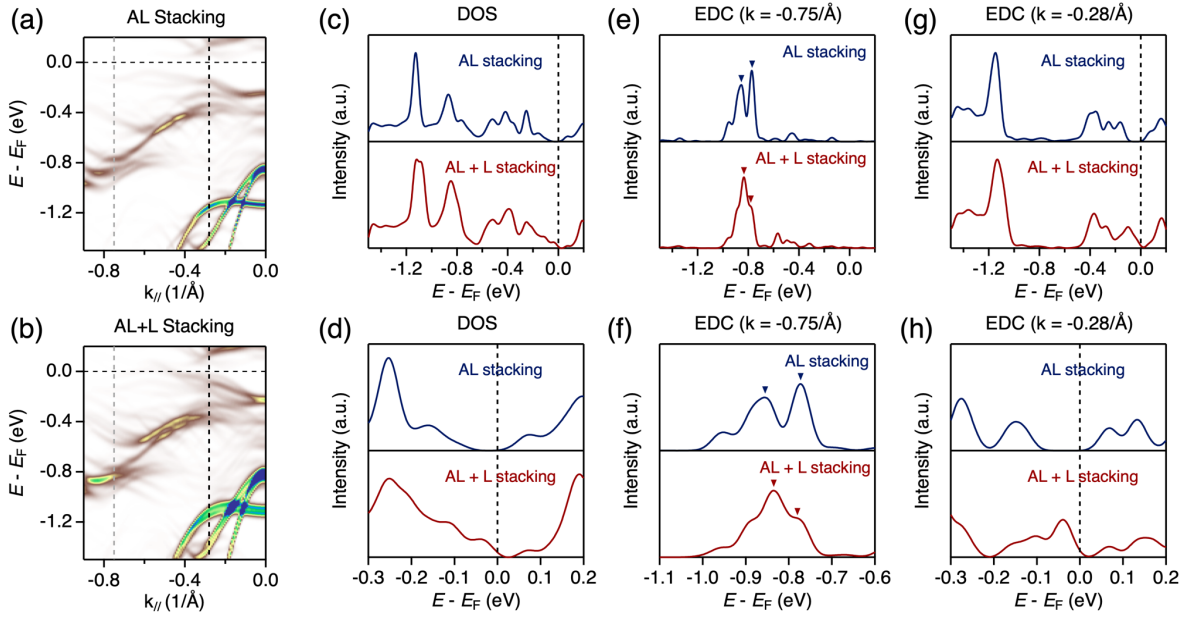

FIG. S9. Theoretical calculations for different interlayer stacking configurations. (a), Calculated band structure along the $\bar{\Gamma} - \bar{M}$ direction for the AL stacking. (b), Similar to panel (a) but for the mixture of AL stacking and L stacking. (c), DOS for the two different stacking configurations in panels (a) and (b), respectively. (d), Zoom-in DOS in the energy range of (-0.3eV, 0.2eV). (e), (f), Similar to panels (c) and (d) but for the EDCs at the position marked by a dashed grey line in panels (a) and (b). (g), (h), Similar to panels (c) and (d) but for the EDCs at the position marked by a dashed black line in panels (a) and (b).

Moreover, Fig. S10 illustrates a comparison between the experimental and calculated results for the hidden state. It is evident that the calculated dispersion of the L stacking configuration exhibits metallic behaviour which is consistent with the observations. However, concerning the CDW gap, it fails to reproduce the gap suppression, but on the contrary, results in an increased gap. Consequently, considering only L stacking cannot fully explain the characteristics of the hidden state. Instead, a combination of both AL stacking and L stacking is essential to accurately depict the features of the hidden state.

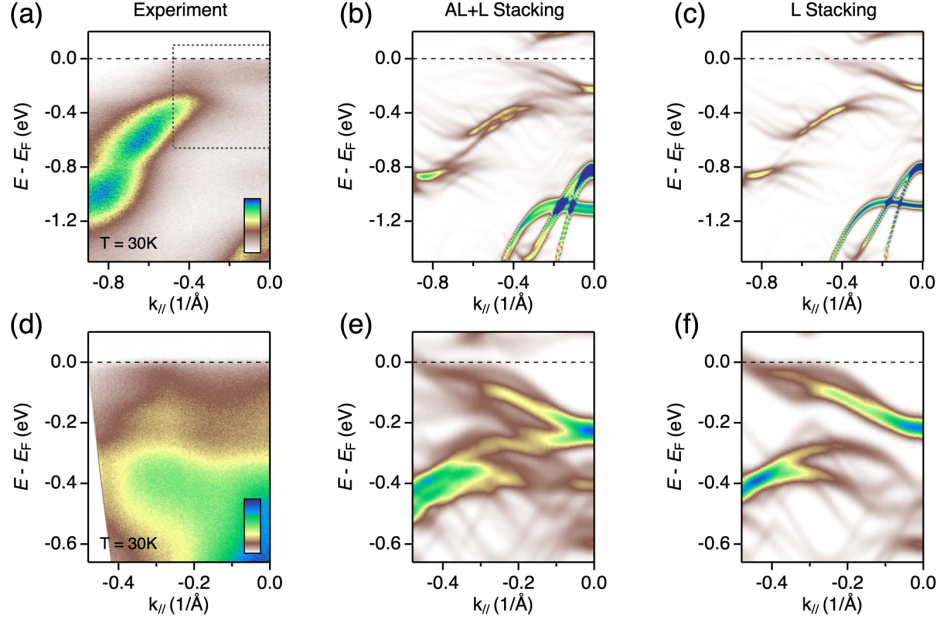

FIG. S10. Comparison of measured hidden state with calculated dispersion. (a), Experimental band structure of the hidden state along the $\bar{\Gamma} - \bar{M}$ direction measured by a helium lamp source (21.2 eV) at T = 30 K. (b), Calculated band structure along the $\bar{\Gamma} - \bar{M}$ direction for a mixture of AL stacking and L stacking. (c), Similar to panel (b) but for the L stacking. (d)-(f), Similar to panels (a)-(c) but for a zoom-in region indicated by a dashed rectangle in panel a, and the experimental band structures are measured by a laser source (7.2 eV).

In addition, examining the evolution of the band structures during the erasing process provides a clearer depiction of changes in interlayer stacking ordering. Specifically, Figs. S11(e) – S11(h) show the progression of the measured dispersions with increasing the number of erasing sequences. Corresponding schematic diagrams are presented in Figs. S11(a) – S11(d), outlining the transition from the hidden state to the C-CDW state, which is accompanied by a reduction in L stacking and the AL domain coarsening. The insets show the intensity variations against the number of erasing sequences, with corresponding states indicated by arrows and highlighted with solid circles. As shown in the calculated dispersions (Figs. S11(i) – S11(l)), a reduction in the proportion of L stacking leads to a gradual opening of the CDW gap and the weakening of the metallic band dispersions, which is in good agreement with the experimental results. Consequently, it is suggested that the hidden state is a mixture of AL stacking and L stacking. Throughout the erasing process, as the material reverts to the C-CDW state, it traverses various intermediate states characterized by the reduction of L stacking and the re-establishment of the long-range interlayer ordering.

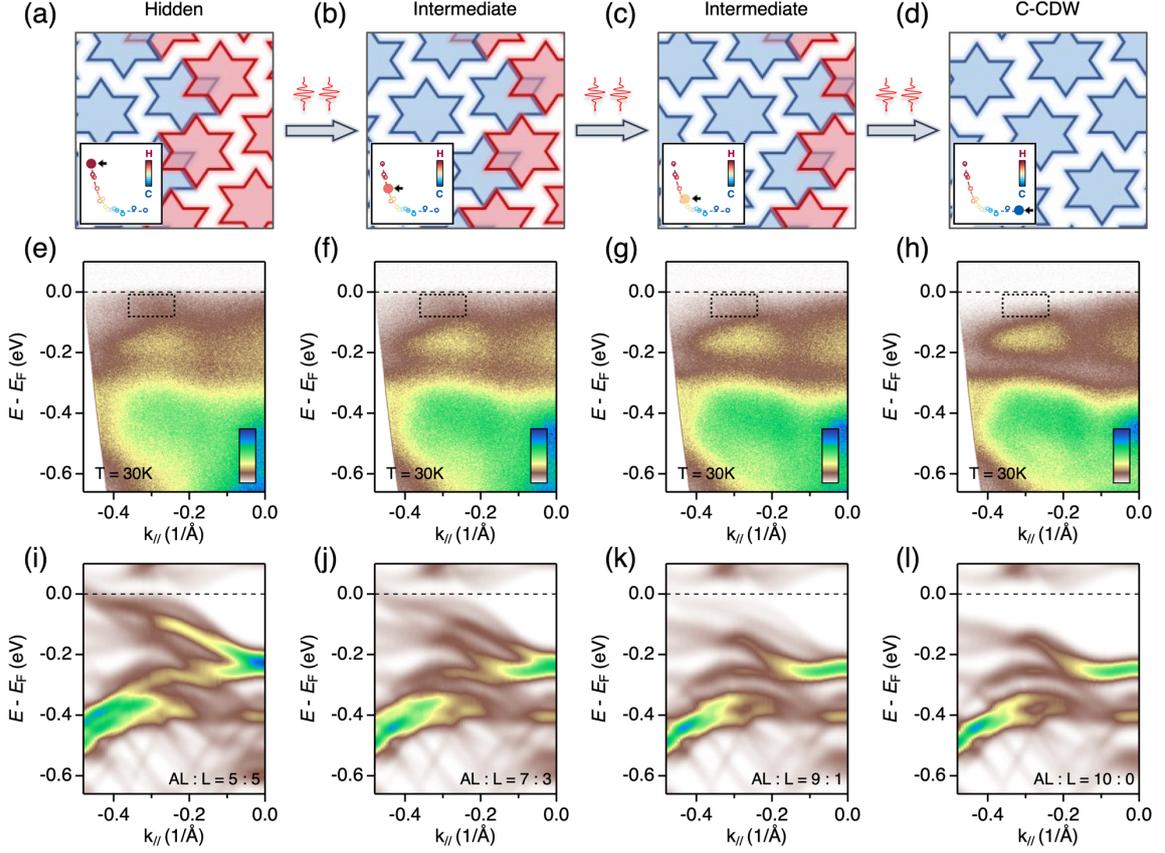

FIG. S11. Evolution of band structures during the continuous erasing process. (a), Schematic illustration of the continuous erasing process from the hidden state (mixture of AL stacking and L stacking) to the C-CDW state (AL stacking). The blue stars represent the AL stacking, and the red stars represent the L stacking. The insets illustrate the intensity of the region marked by a dashed rectangle in panels (e)-(h) as a function of erasing sequence number. The corresponding state are indicated by arrows and highlighted with solid circles. (e)-(h), Corresponding evolution of band structures at different states from the hidden state to the C-CDW state along the $\bar{\Gamma} - \bar{M}$ direction measured by a laser source (7.2 eV) at T = 30 K. (i)-(l), Corresponding evolution of calculated band structures along the $\bar{\Gamma} - \bar{M}$ direction from the hidden state to the C-CDW state with an increasing proportion of AL stacking.